\documentclass[12pt]{article}
\usepackage{times}
\usepackage{geometry}
\usepackage{aas_macros}
\usepackage{xcolor}
\usepackage{comment}
\usepackage[utf8]{inputenc}
\usepackage[utf8]{inputenc}
\usepackage{enumitem,amssymb}
\usepackage{ragged2e}
\newlist{thematic}{itemize}{8}
\setlist[thematic]{label=$\square$}
\usepackage{pifont}
\usepackage{cite}
\usepackage{hyperref}
\usepackage{soul}
\usepackage{multicol}
\usepackage{graphicx}
\usepackage{ulem}
\setlist{nolistsep}

\usepackage{dcolumn}
\usepackage{longtable}
\usepackage{changepage} 
\usepackage{url} 

\begin{document}
\pagenumbering{gobble}


\begin{flushleft}
\huge
 Research and Development for HI Intensity Mapping\linebreak

\normalsize

\noindent \textbf{Thematic Areas:}  
Technological Development Activity, Ground Based Project \linebreak
  
\textbf{Primary Contact:}

Name: Peter Timbie
 \linebreak						
Institution: UW-Madison
 \linebreak
Email: \texttt{pttimbie@wisc.edu}
 \linebreak
Phone:  608-890-2002
 \linebreak
 
 

\vspace*{0.5cm}

\end{flushleft}

\noindent\textbf{Abstract:}
Development of the hardware, data analysis, and simulation techniques for large compact radio arrays dedicated to mapping the $21\,\textrm{cm}$ line of neutral hydrogen gas has proven to be more difficult than imagined twenty years ago when such telescopes were first proposed.  Despite tremendous technical and methodological advances, there are several outstanding questions on how to optimally calibrate and analyze such data. On the positive side, it has become clear that the outstanding issues are purely technical in nature and can be solved with sufficient development activity. Such activity will enable science across redshifts, from early galaxy evolution in the pre-reionization era to dark energy evolution at low redshift.


\vspace*{0.5cm}



\newcommand{\Amherst}{University of Massachusetts, Amherst, MA 01003 USA}
\newcommand{\ANLHEP}{HEP Division, Argonne National Laboratory, Lemont, IL 60439, USA}
\newcommand{\APC}{Laboratoire Astroparticule et Cosmologie (APC), CNRS/IN2P3, Universit\'e Paris Diderot, 10, rue Alice Domon et Léonie Duquet, 75205 Paris Cedex 13, France}
\newcommand{\ASU}{Arizona State University, Tempe, AZ  85287}
\newcommand{\BenGurion}{Department of Physics, Ben-Gurion University, Be'er Sheva 84105, Israel}
\newcommand{\BNL}{Brookhaven National Laboratory, Upton, NY 11973}
\newcommand{\Brown}{Brown University, Providence, RI 02912}
\newcommand{\Bub}{Boston University, Boston, MA 02215}
\newcommand{\BU}{Boston University, Boston, MA 02215}
\newcommand{\Buffalo}{Department of Physics, University at Buffalo, SUNY Buffalo, NY 14260 USA}
\newcommand{\Caltech}{California Institute of Technology, Pasadena, CA 91125}
\newcommand{\Cardiff}{School of Physics and Astronomy, Cardiff University, The Parade, Cardiff, CF24 3AA, UK}
\newcommand{\Carleton}{Carleton University, K1S 5B6 Ottawa, Canada}
\newcommand{\Carnegie}{The Observatories of the Carnegie Institution for Science, 813 Santa Barbara St., Pasadena, CA 91101, USA}
\newcommand{\Cavendish}{Astrophysics Group, Cavendish Laboratory, J.J.Thomson Avenue, Cambridge, CB3 0HE, UK}
\newcommand{\CCA}{Center for Computational Astrophysics, 162 5th Ave, 10010, New York, NY, USA}
\newcommand{\CPPM}{Aix Marseille Univ, CNRS/IN2P3, CPPM, Marseille, France}
\newcommand{\CEADAP}{D\'epartement d’Astrophysique, CEA Saclay DSM/Irfu, 91191 Gif-sur-Yvette, France}
\newcommand{\CERN}{CERN, Geneva, Switzerland}
\newcommand{\CfA}{Harvard-Smithsonian Center for Astrophysics, MA 02138}
\newcommand{\CFT}{Center for Theoretical Physics, Polish Academy of Sciences, al. Lotnik\'{o}w 32/46, 02-668, Warsaw, Poland}
\newcommand{\Cincinnati}{University of Cincinnati, Cincinnati, OH 45221}
\newcommand{\CITA}{Canadian Institute for Theoretical Astrophysics, University of Toronto, Toronto, ON M5S 3H8, Canada}
\newcommand{\CNRSA}{CNRS, Laboratoire d'Annecy-le-Vieux de Physique Th\'{e}orique, France}
\newcommand{\CNYang}{C.N. Yang Institute for Theoretical Physics State University of New York Stony Brook, NY 11794}
\newcommand{\CMUCosmo}{Department of Physics, McWilliams Center for Cosmology, Carnegie Mellon University}
\newcommand{\Columbia}{Columbia University, New York, NY 10027}
\newcommand{\Cornell}{Cornell University, Ithaca, NY 14853}
\newcommand{\CPthree}{CP3-Origins, 5230 Odense, Denmark}
\newcommand{\CUBoulder}{Center for Astrophysics and Space Astronomy, Department of Astrophysical and Planetary Science, University of Colorado, Boulder, CO 80309, USA}
\newcommand{\CWRU}{Case Western Reserve University, Cleveland, OH 44106}
\newcommand{\daa}{Department of Astronomy and Astrophysics, University of Toronto, ON, M5S3H4}
\newcommand{\damtp}{DAMTP, Centre for Mathematical Sciences, Wilberforce Road, Cambridge, UK, CB3 0WA}
\newcommand{\DESY}{DESY,  22607 Hamburg, Germany}
\newcommand{\dfa}{Departamento de F\'{\i}sica e Astronomia, Faculdade de Ci\^{e}ncias, Universidade do Porto, Porto, Portugal}
\newcommand{\DFI}{Departamento de F\'isica, FCFM, Universidad de Chile, Blanco Encalada 2008, Santiago, Chile}
\newcommand{\DOE}{US. Department of Energy, Germantown, MD 20874}
\newcommand{\drexel}{Drexel University, Philadelphia, PA 19104}
\newcommand{\Duke}{Duke University and Triangle Universitites Nuclear Laboratory, Durham, NC 27708}
\newcommand{\DukePhys}{Department of Physics, Duke University, Durham, NC 27708, USA}
\newcommand{\dunlap}{Dunlap Institute for Astronomy and Astrophysics, University of Toronto, ON, M5S3H4}
\newcommand{\Durham}{Department of Physics, Lower Mountjoy, South Rd, Durham DH1 3LE, United Kingdom}
\newcommand{\ED}{University of Edinburgh, EH8 9YL Edinburgh, United Kingdom}
\newcommand{\EPFL}{Institute of Physics, Laboratory of Astrophysics, Ecole Polytechnique Fédérale de Lausanne (EPFL), Observatoire de Sauverny, 1290 Versoix, Switzerland}
\newcommand{\ETH}{ETH Zurich, Institute for Particle Physics, 8093 Zurich, Switzerland}
\newcommand{\FNAL}{Fermi National Accelerator Laboratory, Batavia, IL 60510}
\newcommand{\FQAUB}{Dept. de F\' isica Qu\` antica i Astrof\' isica, Universitat de Barcelona, Mart\' i i Franqu\` es 1, E08028 Barcelona, Spain}
\newcommand{\FSU}{Florida State University, Tallahassee, FL 32306}
\newcommand{\Glasgow}{University of Glasgow, G12 8QQ Glasgow, United Kingdom}
\newcommand{\GRAPPA}{GRAPPA Institute, University of Amsterdam, Science Park 904, 1098 XH Amsterdam, The Netherlands}
\newcommand{\GSFC}{Goddard Space Flight Center, Greenbelt, MD 20771 USA}
\newcommand{\GWU}{George Washington University, Washington, DC 20052}
\newcommand{\Hampton}{Hampton University, Hampton, VA 23668}
\newcommand{\HarvardPhys}{Department of Physics, Harvard University, Cambridge, MA 02138, USA}
\newcommand{\Haverford}{Haverford College, 370 Lancaster Ave, Haverford PA, 19041, USA}
\newcommand{\Hawaii}{University of Hawaii, Honolulu, HI 96822}
\newcommand{\HKUST}{The Hong Kong University of Science and Technology, Hong Kong SAR, China}
\newcommand{\houston}{University of Houston, Houston, TX 77204}
\newcommand{\IA}{Instituto de Astrof\'{\i}sica e Ci\^encias do Espa\c{c}o (IA), Porto, Portugal}
\newcommand{\IAC}{Instituto de Astrof \'{\i}isica de Canarias, 38200 La Laguna, Tenerife, Spain}
\newcommand{\IAP}{Institut d'Astrophysique de Paris (IAP), CNRS \& Sorbonne University, Paris, France}
\newcommand{\IAS}{Institute for Advanced Study, Princeton, NJ 08540}
\newcommand{\IBS}{Institute for Basic Science (IBS), Daejeon 34051, Korea}
\newcommand{\ICC}{ICC, University of Barcelona, IEEC-UB, Mart\' i i Franqu\` es, 1, E08028 Barcelona, Spain}
\newcommand{\ICCD}{Institute for Computational Cosmology, Department of Physics, Durham University, South Road, Durham, DH1 3LE, UK}
\newcommand{\ICE}{Institute of Space Sciences (ICE, CSIC), Campus UAB, Carrer de Can Magrans, s/n, 08193 Barcelona, Spain}
\newcommand{\ICRR}{Institute for Cosmic Ray Resaerch, The University of Tokyo, 456 Higashi-Mozumi, Kamioka, Hida, Gifu 506-1205, Japan}
\newcommand{\ICTP}{International Centre for Theoretical Physics, Strada Costiera, 11, I-34151 Trieste, Italy}
\newcommand{\IFAE}{Institut de Fisica d’Altes Energies, The Barcelona Institute of Science and Technology, Campus UAB, 08193 Bellaterra (Barcelona), Spain}
\newcommand{\IFPU}{IFPU - Institute for Fundamental Physics of the Universe, Via Beirut 2, 34014 Trieste, Italy}
\newcommand{\IFT}{Instituto de Fisica Teorica UAM/CSIC, Universidad Autonoma de Madrid, 28049 Madrid, Spain}
\newcommand{\IFUNAM}{IFUNAM - Instituto de F\'{i}sica, Universidad Nacional Aut\'onoma de M\'etico, 04510 CDMX, M\'exico}
\newcommand{\IHEP}{Institute of High Energy Physics, Austrian Academy of Sciences, 1050 Vienna, Austria}
\newcommand{\Imperial}{Theoretical Physics, Blackett Laboratory, Imperial College, London, SW7 2AZ, U.K.}
\newcommand{\Indiana}{Indiana University, Bloomington, IN 47405}
\newcommand{\INAFOATs}{INAF - Osservatorio Astronomico di Trieste, Via G.B. Tiepolo 11, 34143 Trieste, Italy}
\newcommand{\INAFOAS}{INAF - Osservatorio di Astrofisica e Scienza dello Spazio di Bologna, via Piero Gobetti 93/3, I-40129 Bologna, Italy}
\newcommand{\INFNCag}{Istituto Nazionale di Fisica Nucleare, Sezione di Cagliari,  09126 Cagliari, Italy}
\newcommand{\INFNCat}{Istituto Nazionale di Fisica Nucleare, Sezione di Catania, 95125 Catania, Italy}
\newcommand{\INFNG}{Istituto Nazionale di Fisica Nucleare, Sezione di Genova, 16146 Genova, Italy}
\newcommand{\INFN}{INFN – National Institute for Nuclear Physics, Via Valerio 2, I-34127 Trieste, Italy}
\newcommand{\INFNFE}{Istituto Nazionale di Fisica Nucleare, Sezione di Ferrara, 40122, Italy }
\newcommand{\INFNLNF}{Istituto Nazionale di Fisica Nucleare, Laboratori Nazionali di Frascati, 00044 Frascati, Italy}
\newcommand{\INFNLNS}{Istituto Nazionale di Fisica Nucleare, Laboratori Nazionali del Sud, 95125 Catania, Italy}
\newcommand{\INFNN}{Istituto Nazionale di Fisica Nucleare, Sezione di Napoli, 80125 Napoli, Italy }
\newcommand{\INFNPD}{Istituto Nazionale di Fisica Nucleare, Sezione di Padova, via Marzolo 8, I-35131, Padova, Italy}
\newcommand{\INFNRM}{Istituto Nazionale di Fisica Nucleare, Sezione di Roma, 00185 Roma, Italy}
\newcommand{\INFNT}{Istituto Nazionale di Fisica Nucleare, Sezione di Torino, 10125, Italy }
\newcommand{\ioa}{Institute of Astronomy, University of Cambridge,Cambridge CB3 0HA, UK}
\newcommand{\IPP}{Institute for Particle Physics, BC V8W 3P6 Victoria, Canada}
\newcommand{\IPMU}{Kavli Insitute for the Physics and Mathematics of the Universe (WPI), University of Tokyo, 277-8583 Kashiwa , Japan}
\newcommand{\IPNL}{Universit\'e de Lyon, F-69622, Lyon, France; Universit\'e de Lyon 1, Villeurbanne; CNRS/IN2P3, Institut de Physique Nucl\'eaire de Lyon}
\newcommand{\IRFU}{IRFU, CEA, Universit\'e Paris-Saclay, F-91191 Gif-sur-Yvette, France}
\newcommand{\ITFA}{Institute for Theoretical Physics, University of Amsterdam, Science Park 904, 1098 XH Amsterdam, The Netherlands}
\newcommand{\IUCAA}{The Inter-University Centre for Astronomy and Astrophysics, Pune, 411007, India}
\newcommand{\Jerusalem}{Hebrew University of Jerusalem, 91904 Jerusalem, Israel}
\newcommand{\JHU}{Johns Hopkins University, Baltimore, MD 21218}
\newcommand{\JLAB}{Thomas Jefferson National Laboratory, Newport News, VA 23606}
\newcommand{\JPL}{Jet Propulsion Laboratory, California Institute of Technology, Pasadena, CA, USA}
\newcommand{\KASSI}{Korea Astronomy and Space Science Institute, Daejeon 34055, Korea}
\newcommand{\kavli}{Kavli Institute for Cosmology, Cambridge, UK, CB3 0HA}
\newcommand{\KIAS}{School of Physics, Korea Institute for Advanced Study, 85 Hoegiro, Dongdaemun-gu, Seoul 130-722, Korea}
\newcommand{\KICP}{Kavli Institute for Cosmological Physics, Chicago, IL 60637}
\newcommand{\KIPAC}{Kavli Institute for Particle Astrophysics and Cosmology, Stanford 94305}
\newcommand{\KINGS}{King's College London, WC2R 2LS London, United Kingdom}
\newcommand{\Kobe}{Kobe University, 657-8501 Kobe, Japan}
\newcommand{\KPH}{Johannes Gutenberg University, 55128 Mainz, Germany}
\newcommand{\KPMU}{University of Tokyo, 277-8583  Kashiwa , Japan}
\newcommand{\KSU}{Kansas State University, Manhattan, KS 66506}
\newcommand{\KwaZuluNatal}{Astrophysics and Cosmology Research Unit, School of Chemistry and Physics, University of KwaZulu-Natal, Durban 4000, South Africa}
\newcommand{\Lafayette}{Lafayette College, Easton, PA 18042}
\newcommand{\LANL}{Los Alamos National Laboratory, Los Alamos, NM 87545}
\newcommand{\LBL}{Lawrence Berkeley National Laboratory, Berkeley, CA 94720}
\newcommand{\Leiden}{Lorentz Institute, Leiden University, Niels Bohrweg 2,Leiden, NL 2333 CA, The Netherlands}
\newcommand{\Liverpool}{University of Liverpool,  L69 7ZE Liverpool , United Kingdom}
\newcommand{\LLNL}{Lawrence Livermore National Laboratory, Livermore, CA, 94550}
\newcommand{\LPC}{Universit\'e Clermont Auvergne, CNRS/IN2P3, Laboratoire de Physique de Clermont, F-63000 Clermont-Ferrand, France}
\newcommand{\LPNHE}{Sorbonne Universit\'e, Universit\'e Paris Diderot, CNRS/IN2P3, Laboratoire de Physique Nucl\'eaire et de Hautes Energies, LPNHE, 4 Place Jussieu, F-75252 Paris, France}
\newcommand{\McGill}{McGill University, Montreal, QC H3A 2T8, Canada}
\newcommand{\Melbourne}{School of Physics, The University of Melbourne, Parkville, VIC 3010, Australia}
\newcommand{\Mines}{Colorado School of Mines, Golden, CO 80401}
\newcommand{\MIT}{Massachusetts Institute of Technology, Cambridge, MA 02139}
\newcommand{\MPE}{Max-Planck-Institut f\"{u}r extraterrestrische Physik (MPE), Giessenbachstrasse 1, D-85748 Garching bei M\"unchen, Germany}
\newcommand{\MPIA}{Max-Planck-Institut f\"{u}r Astrophysik, Karl-Schwarzschild-Str. 1, 85741 Garching, Germany}
\newcommand{\MPP}{Max-Planck-Institut f\"{u}r Physik (Werner-Heisenberg-Institut), F\"ohringer Ring 6, D-80805 M\"unchen, Germany}
\newcommand{\LUPM}{Laboratoire Univers et Particules de Montpellier, Univ. Montpellier and CNRS, 34090 Montpellier, France}
\newcommand{\NAOC}{National Astronomical Observatories, Chinese Academy of Sciences, PR China}
\newcommand{\NCBJ}{National Center for Nuclear Research, Ul.Pasteura 7,Warsaw, Poland}
\newcommand{\NCU}{National Central University, Taoyuan City 32001, Taiwan (R.O.C.)}
\newcommand{\NCSU}{Physics Department, North Carolina State Universitym, 2401 Stinson Dr, Raleigh, NC 27607}
\newcommand{\ND}{University of Notre Dame,vNotre Dame, IN 46556}
\newcommand{\NIU}{Northern Illinois University, DeKalb, Illinois 60115}
\newcommand{\NMSU}{New Mexico State University, Las Cruces, NM 88003}
\newcommand{\NOAO}{National Optical Astronomy Observatory, 950 N. Cherry Ave., Tucson, AZ 85719 USA}
\newcommand{\Northwestern}{Northwestern University, Evanston, IL 60201}
\newcommand{\Nottingham}{University of Nottingham, NG7 2RD Nottingham, United Kingdom}
\newcommand{\NPPSFAmes}{NASA Postdoctoral Program Senior Fellow, NASA Ames Research Center, Moffett Field, CA 94035, USA}
\newcommand{\NWU}{Northwestern University, Evanston, IL 60208}
\newcommand{\NYU}{New York University, New York, NY 10003}
\newcommand{\OK}{ University of Oklahoma, Norman, OK 73019}
\newcommand{\ORNL}{Oak Ridge National Laboratory, Oak Ridge, TN 37831}
\newcommand{\OSU}{The Ohio State University, Columbus, OH 43212}
\newcommand{\OU}{Department of Physics and Astronomy, Ohio University, Clippinger Labs, Athens, OH 45701, USA}
\newcommand{\OskarKlein}{Oskar Klein Centre for Cosmoparticle Physics, Stockholm University, AlbaNova, Stockholm SE-106 91, Sweden}
\newcommand{\Oxford}{The University of Oxford, Oxford OX1 3RH, UK}
\newcommand{\Oxy}{Occidental College, Los Angeles, CA 90041}
\newcommand{\ParisSud}{Universit\'{e} Paris-Sud, LAL, UMR 8607, F-91898 Orsay Cedex, France \& CNRS/IN2P3, F-91405 Orsay, France}
\newcommand{\PI}{Perimeter Institute, Waterloo, Ontario N2L 2Y5, Canada}
\newcommand{\Pitt}{University of Pittsburgh and PITT PACC, Pittsburgh, PA 15260}
\newcommand{\PNNL}{Pacific Northwest National Laboratory ,Richland, WA 99352}
\newcommand{\PNPI}{Petersburg Nuclear Physics Institute, 188300 Gatchina, Russia}
\newcommand{\Port}{Institute of Cosmology \& Gravitation, University of Portsmouth, Dennis Sciama Building, Burnaby Road, Portsmouth PO1 3FX, UK}
\newcommand{\Princeton}{Princeton University, Princeton, NJ 08544}
\newcommand{\PSU}{The Pennsylvania State University, University Park, PA 16802}
\newcommand{\Purdue}{Purdue University, West Lafayette, IN 47907}
\newcommand{\PW}{Participation Worldscope, Sedona, Arizona and Hong Kong, SAR PRC}
\newcommand{\Queens}{Queen's University , K7L 3N6 Kingston, Canada}
\newcommand{\Queensland}{The University of Queensland, School of Mathematics and Physics, QLD 4072, Australia}
\newcommand{\QMUL}{Queen Mary University of London, Mile End Road, London E1 4NS, United Kingdom}
\newcommand{\RAL}{Radio Astronomy Laboratory, University of California Berkeley, Berkeley, CA 94720, USA}
\newcommand{\Rice}{Department of Physics \& Astronomy, Rice University, Houston, Texas 77005, USA}
\newcommand{\RIT}{Rochester Institute of Technology}
\newcommand{\RomaS}{Dipartimento di Fisica, Universit\`{a} La Sapienza, P. le A. Moro 2, Roma, Italy}
\newcommand{\RUG}{Kapteyn Astronomical Institute, University of Groningen, P.O. Box 800, 9700 AV Groningen, The Netherlands}
\newcommand{\Rutgers}{Department of Physics and Astronomy, Rutgers, the State University of New Jersey, 136 Frelinghuysen Road, Piscataway, NJ 08854, USA}
\newcommand{\Sanford}{Sanford Underground Research Facility, Lead, SD 57754}
\newcommand{\Sassari}{Universit\`a di Sassari, 07100 Sassari,  Italy}
\newcommand{\SCIPP}{University of California at Santa Cruz, Santa Cruz, CA 95064}
\newcommand{\Sejong}{Department of Physics and Astronomy, Sejong University, Seoul, 143-747, Korea}
\newcommand{\Sheffield}{University of Sheffield, S3 7RH Sheffield, United Kingdom}
\newcommand{\SHAO}{Shanghai Astronomical Observatory (SHAO), Nandan Road 80, Shanghai 200030, China}
\newcommand{\Siena}{Siena College, 515 Loudon Road, Loudonville, NY 12211, USA}
\newcommand{\SISSA}{SISSA - International School for Advanced Studies, Via Bonomea 265, 34136 Trieste, Italy}
\newcommand{\SimonFraser}{Department of Physics, Simon Fraser University, Burnaby, British Columbia, Canada V5A 1S6}
\newcommand{\SLAC}{SLAC National Accelerator Laboratory, Menlo Park, CA 94025}
\newcommand{\SMU}{Southern Methodist University, Dallas, TX 75275}
\newcommand{\SNOLAB}{SNOLAB, Lively, ON P3Y 1N2, Canada}
\newcommand{\SoCal}{University of Southern California, CA 90089 }
\newcommand{\Stanford}{Stanford University, Stanford, CA 94305}
\newcommand{\StonyBrook}{Stony Brook University, Stony Brook, NY 11794}
\newcommand{\STSCI}{Space Telescope Science Institute, Baltimore, MD 21218}
\newcommand{\SUNYA}{University at Albany SUNY, Albany, NY 12222}
\newcommand{\SussexAstronomy}{Astronomy Centre, School of Mathematical and Physical Sciences, University of Sussex, Brighton BN1 9QH, United Kingdom}
\newcommand{\Syracuse}{Syracuse University, Syracuse, NY 13244}
\newcommand{\Tamu}{Texas AandM University, College Station, TX 77843 }
\newcommand{\Techsource}{Techsource Incorporated, Los Alamos, NM 87544}
\newcommand{\TelAviv}{Tel-Aviv University,  69978 Tel-Aviv, Israel}
\newcommand{\Temple}{Temple University, Philadelphia, PA 19122}
\newcommand{\TIFR}{Tata Institute of Fundamental Research, Homi Bhabha Road, Mumbai 400005 India}
\newcommand{\Tsinghua}{Department of Physics and Tsinghua Center for Astrophysics, Tsinghua University, Beijing 100084, P R China}
\newcommand{\TUM}{Technical University of Munich,  80333 Munich, Germany}
\newcommand{\UA}{University of Alabama, Tuscaloosa, AL 35487}
\newcommand{\UAS}{Department of Astronomy/Steward Observatory, University of Arizona, Tucson, AZ  85721}
\newcommand{\UAM}{Universidad Aut\'onoma de Madrid, 28049, Madrid, Spain}
\newcommand{\UBC}{University of British Columbia, Vancouver, BC V6T 1Z1, Canada}
\newcommand{\UCB}{Department of Astronomy, University of California Berkeley, Berkeley, CA 94720, USA}
\newcommand{\UCBP}{Department of Physics, University of California Berkeley, Berkeley, CA 94720, USA}
\newcommand{\UCBSSL}{Space Sciences Laboratory, University of California Berkeley, Berkeley, CA 94720, USA}
\newcommand{\UCD}{University of California at Davis, Davis, CA 95616}
\newcommand{\UChicago}{University of Chicago, Chicago, IL 60637}
\newcommand{\UCI}{University of California, Irvine, CA 92697}
\newcommand{\UCLA}{University of California at Los Angeles, Los Angeles,  CA 90095}
\newcommand{\UCL}{University College London, WC1E 6BT London, United Kingdom}
\newcommand{\UCR}{University of California at Riverside, Riverside, CA 92521}
\newcommand{\UCSB}{University of California at Santa Barbara, Santa Barbara, CA 93106}
\newcommand{\UCSC}{University of California at Santa Cruz, Santa Cruz, CA 95064}
\newcommand{\UCSD}{University of California San Diego, La Jolla, CA 92093}
\newcommand{\UFL}{University of Florida, Gainesville, FL 32611}
\newcommand{\UFN}{Universit\`a Federico II di Napoli, 80125 Napoli, Italy}
\newcommand{\UGTO}{Divisi\'on de Ciencias e Ingenier\'ias, Universidad de Guanajuato, Le\'on 37150, M\'exico}
\newcommand{\UKY}{University of Kentucky, Lexington, KY 40506}
\newcommand{\UMD}{University of Maryland, College Park, MD 20742
\newcommand{\UMiami}{University of Miami, Coral Gables, FL 33124}}
\newcommand{\UMich}{University of Michigan, Ann Arbor, MI 48109}
\newcommand{\UMN}{University of Minnesota, Minneapolis, MN 55455}
\newcommand{\UnB}{Instituto de F\'{i}sica, Universidade de Bras\'{i}lia, 70919-970, Bras\'{i}lia, DF, Brazil}
\newcommand{\UNC}{University of North Carolina at Chapel Hill, Chapel Hill, NC 27599}
\newcommand{\UNH}{University of New Hampshire, Durham, NH 03824}
\newcommand{\UNIMI}{Dipartimento di Fisica ``Aldo Pontremoli'', Universit\`a{} degli Studi di Milano, via Celoria 16, 20133 Milano, Italy}
\newcommand{\UNIPD}{Dipartimento di Fisica e Astronomia ``G. Galilei'',Universit\`a degli Studi di Padova, via Marzolo 8, I-35131, Padova, Italy}
\newcommand{\UNM}{University of New Mexico, Albuquerque, NM 87131}
\newcommand{\UNV}{University of Nevada, Reno, NV 89557}
\newcommand{\UoM}{Jodrell Bank Center for Astrophysics, School of Physics and Astronomy, University of Manchester, Oxford Road, Manchester, M13 9PL, UK}
\newcommand{\UPenn}{Department of Physics and Astronomy, University of Pennsylvania, Philadelphia, Pennsylvania 19104, USA}
\newcommand{\UR}{Department of Physics and Astronomy, University of Rochester, 500 Joseph C. Wilson Boulevard, Rochester, NY 14627, USA}
\newcommand{\UrbanaC}{Department of Physics, University of Illinois at Urbana-Champaign, Urbana, Illinois 61801, USA}
\newcommand{\USC}{The University of South Carolina, Columbia, SC 29208}
\newcommand{\USD}{The University of South Dakota, Vermillion, SD 57069}
\newcommand{\UTD}{University of Texas at Dallas, Texas 75080}
\newcommand{\Utenn}{The University of Tennessee, Knoxville, TN 37996}
\newcommand{\Utah}{University of Utah, Department of Physics and Astronomy, 115 S 1400 E, Salt Lake City, UT 84112, USA}
\newcommand{\UVA}{University of Virginia, Charlottesville, VA 22903}
\newcommand{\Uvic}{University of Victoria, BC V8P 5C2 Victoria, Canada}
\newcommand{\UWaterloo}{Department of Physics and Astronomy, University of Waterloo, 200 University Ave W, Waterloo, ON N2L 3G1, Canada}
\newcommand{\UWMadison}{Department of Physics, University of Wisconsin - Madison, Madison, WI 53706}
\newcommand{\UW}{University of Washington, Seattle 98195}
\newcommand{\UWC}{Department of Physics \& Astronomy, University of the Western Cape, Cape Town 7535, South Africa}
\newcommand{\Vanderbilt}{Physics \& Astronomy Department, Vanderbilt University, PMB 401807, 2301 Vanderbilt Place, Nashville, TN 37235}
\newcommand{\VSI}{Van Swinderen Institute for Particle Physics and Gravity, University of Groningen, Nijenborgh 4, 9747~AG~Groningen, The~Netherlands}
\newcommand{\VT}{Virginia Tech, Blacksburg, VA 24061}
\newcommand{\VUU}{Virginia Union University, Richmond, Virginia, 23220}
\newcommand{\WCA}{Centre for Astrophysics, University of Waterloo, Waterloo, Ontario N2L 3G1, Canada}
\newcommand{\Weizmann}{Weizmann Institute of Science, 76100 Rehovot, Israel}
\newcommand{\Wellesley}{Wellesley College, Wellesley, MA 02481}
\newcommand{\wiscIce}{University of Wisconsin, Madison, WI 53706}
\newcommand{\WM}{College of William and Mary, Newport News, VA 23606}
\newcommand{\WUSL}{Washington University in St Louis, St. Louis, MO 63130}
\newcommand{\WVU}{CSEE, West Virginia University, Morgantown, WV 26505, USA}
\newcommand{\WVUGWAC}{Center for Gravitational Waves and Cosmology, West Virginia University, Morgantown, WV 26505, USA}
\newcommand{\Wyoming}{Department of Physics and Astronomy, University of Wyoming, Laramie, WY 82071, USA}
\newcommand{\Yale}{Department of Physics, Yale University, New Haven, CT 06520}
\newcommand{\YorkU}{Department of Physics and Astronomy, York University, Toronto, Ontario M3J 1P3, Canada}

\noindent\textbf{Contributors and Endorsers:} 
\noindent Zeeshan Ahmed$^{1}$, 
David Alonso$^{2}$, 
Mustafa A. Amin$^{3}$, 
R\'{e}za Ansari$^{4}$, 
Evan J. Arena$^{5,6}$, 
Kevin Bandura$^{7,8}$, 
Adam Beardsley$^{9}$, 
Philip Bull$^{10,11}$, 
Emanuele Castorina$^{12}$, 
Tzu-Ching Chang$^{13}$, 
Romeel Dav\'e$^{14}$, 
Alexander van Engelen$^{15,9}$, 
Aaron Ewall-Wice$^{13}$, 
Simone Ferraro$^{16}$, 
Simon Foreman$^{15}$, 
Josef Frisch$^{1}$, 
Daniel Green$^{17}$, 
Gilbert Holder$^{18}$, 
Daniel Jacobs$^{9}$, 
Joshua~S.~Dillon$^{19}$, 
Dionysios Karagiannis$^{20,21}$, 
Alexander A. Kaurov$^{22}$, 
Lloyd Knox$^{23}$, 
Emily Kuhn$^{24}$, 
Adrian Liu$^{25}$, 
Yin-Zhe Ma$^{26}$, 
Kiyoshi W. Masui$^{27}$, 
Thomas McClintock$^{5}$, 
Kavilan Moodley$^{26}$, 
Moritz M{\"u}nchmeyer$^{28}$,
Laura B. Newburgh$^{24}$, 
Andrei Nomerotski$^{5}$, 
Paul O'Connor$^{5}$, 
Andrej Obuljen$^{29}$, 
Hamsa Padmanabhan$^{15}$, 
David Parkinson$^{30}$, 
Olivier Perdereau$^{4}$, 
David Rapetti$^{31,32}$, 
Benjamin Saliwanchik$^{24}$, 
Neelima Sehgal$^{33}$, 
J. Richard Shaw$^{34}$, 
Chris Sheehy$^{5}$, 
Erin Sheldon$^{5}$, 
Raphael Shirley$^{35}$, 
Eva Silverstein$^{36}$, 
Tracy Slatyer$^{27,22}$, 
An\v{z}e Slosar$^{5}$, 
Paul Stankus$^{5}$, 
Albert Stebbins$^{37}$, 
Peter T. Timbie$^{38}$, 
Gregory S. Tucker$^{39}$, 
William Tyndall$^{5,24}$, 
Francisco Villaescusa-Navarro$^{40}$, 
Dallas Wulf$^{25}$

\vspace*{0.4cm}

{\small
\noindent$^{1}$ \SLAC \\
$^{2}$ \Oxford \\
$^{3}$ \Rice \\
$^{4}$ \ParisSud \\
$^{5}$ \BNL \\
$^{6}$ \drexel \\
$^{7}$ \WVU \\
$^{8}$ \WVUGWAC \\
$^{9}$ \ASU \\
$^{10}$ \QMUL \\
$^{11}$ \UWC \\
$^{12}$ \UCBP \\
$^{13}$ \JPL \\
$^{14}$ \ED \\
$^{15}$ \CITA \\
$^{16}$ \LBL \\
$^{17}$ \UCSD \\
$^{18}$ \UrbanaC \\
$^{19}$ \UCB \\
$^{20}$ \UNIPD \\
$^{21}$ \INFNPD \\
$^{22}$ \IAS \\
$^{23}$ \UCD \\
$^{24}$ \Yale \\
$^{25}$ \McGill \\
$^{26}$ \KwaZuluNatal \\
$^{27}$ \MIT \\
$^{28}$ \PI \\
$^{29}$ \WCA \\
$^{30}$ \KASSI \\
$^{31}$ \CUBoulder \\
$^{32}$ \NPPSFAmes \\
$^{33}$ \StonyBrook \\
$^{34}$ \UBC \\
$^{35}$ \IAC \\
$^{36}$ \Stanford \\
$^{37}$ \FNAL \\
$^{38}$ \UWMadison \\
$^{39}$ \Brown \\
$^{40}$ \CCA \\
}

\pagebreak
\pagenumbering{arabic}

\setlength{\parskip}{3pt}
\setlength{\parindent}{18pt}
\section{Key Science Goals \& Objectives}
\vspace*{-9pt}

Three-dimensional surveys with the redshifted $21\,\textrm{cm}$ line are key to achieving many of the goals outlined by Astro2020 Science White Papers. In the near future, this technique, called {\it $21\,\textrm{cm}$ intensity mapping}, will likely surpass optical surveys in terms of volume of the cosmos surveyed.  The science goals include:  exploration of the cosmic Dark Ages, Cosmic Dawn and the Epoch of Reionization\cite{Mirocha2019, Furlanetto2019DarkAges,Furlanetto2019EoR,Furlanetto2019Synergy, Hutter2019,Chang2019,Alvarez2019, Cooray2019, Kovetz2019}; understanding inflation\cite{Liu2019Cosmology,Kovetz2019,Meerburg2019,Slosar2019Inflation};  understanding dark energy and modified gravity\cite{Slosar2019DE};  and determining neutrino mass\cite{Dvorkin2019}. In addition, owing to their large fields of view and high survey speeds, 
many $21\,\textrm{cm}$ intensity mapping instruments can simultaneously monitor the sky for transient events, such as pulsars and fast radio bursts (FRBs).  Science white papers\cite{Lynch2019, Ravi2019, Law2019} outline plans to understand the FRB mechanism and use them as cosmological probes. And some intensity mapping instruments will discover new millisecond pulsars, essential for the improved pulsar timing arrays for gravitational wave detection, as described in white papers\cite{Kelley2019, Taylor2019, Cordes2019, Siemens2019, Fonseca2019, Lorimer2019}. 

After the recombination of hydrogen, when the Cosmic Microwave Background (CMB) was
created at redshifts around $z\sim 1150$, the baryonic portion of the universe was dominated by neutral hydrogen.  As matter continued
to cluster in the post-recombination universe, peaks in the matter density
grew and eventually led to the formation of the first generation of
stars and galaxies, which emitted radiation capable of reionizing the
ambient neutral hydrogen.  Between
recombination and the formation of the first stars, during the high-redshift epoch generally referred to as the Dark Ages $(30 \lesssim z \lesssim 150)$, neutral hydrogen traces the overall matter distribution. 
$21\,\textrm{cm}$ intensity mapping is the only known technique for accessing this epoch and could provide large-scale tomographic maps sampling vastly more of the pristine density fluctuation modes than can the CMB.  

Later, between $z\sim30$ and $z\sim6$, first-generation stars
and galaxies formed and began the process of reionizing the
universe. During this epoch, including Cosmic Dawn and the Epoch of Reionization (EoR), 
the signal is boosted by large regions of completely ionized
``bubbles'' residing in a sea of otherwise largely neutral hydrogen. Probing this unexplored phase of cosmic evolution was the top-ranked science program from the Astro2010 Radio, Millimeter, and Submillimeter panel. Indeed, 
a number of experiments, such as 
LWA, HERA, PAPER, LOFAR, MWA, and GMRT, 
are seeking to make the first
measurements of how the first luminous objects affected the
large-scale distribution and ionization state of hydrogen.  They are spurred on by the possible detection of the signature of the formation of the first stars in the global spectrum\cite{2018Natur.555...67B} (as opposed to maps) of the $21\,\textrm{cm}$.
Because CMB photons scatter from the ionized matter, these intensity mapping measurements are critical to interpreting CMB power spectra.  In particular, they can provide independent (of the CMB) measurements of $\tau$, the optical depth of reionization, which is required for CMB constraints on neutrino mass\cite{Liu2016}.  

Finally, in the post-EoR epoch, $z\lesssim 6$, the universe is mostly
ionized, with a few pockets of neutral hydrogen residing in
galaxies. $21\,\textrm{cm}$ intensity mapping can measure the large scale structure in this epoch by surveying the aggregate emission from many unresolved galaxies.  Even without resolving individual objects, one
can still trace the fluctuations in their number density across space
and redshift on large scales,  where theoretical modeling is most robust.
A key goal is to determine the expansion history of the universe spanning redshifts $z=0.3-6$, complementing existing measurements at low redshift while opening up new windows at high redshifts. Another goal is to measure the growth-rate of structure formation over this same wide range of redshift and thus constrain modifications of gravity over a wide range of scales and times in cosmic history.  One can also observe, or constrain, the presence of inflationary relics
in the primordial power spectrum and observe, or constrain, primordial non-Gaussianity.
The first generation of experiments in this redshift range include GBT, CHIME, Tianlai, HIRAX, OWFA, BINGO, BMX, and PAON.  The proposed PUMA experiment\cite{PUMAWhitePaper} targets these science goals over the entire post-EoR epoch.
Many of these instruments also include hardware and software for detecting radio transients.  CHIME, for example, has already detected 13 FRBs\cite{CHIME2019} and is expected to find many more.

Reaching these science goals will require overcoming several challenges, including observing the $21\,\textrm{cm}$ line over an enormous range of frequencies, from $\sim10 - 1400$\,MHz.  However, all current observational programs, from Cosmic Dawn/EoR to post-EoR, have converged on broadly similar approaches.  As outlined below, they require continued development of common hardware, data analysis, and simulation techniques. (A whitepaper focusing on Cosmic Dawn/EoR complements much of the discussion here\cite{AdrianWhitePaper}.)  Lessons learned from these ongoing programs will open wide a new window for astrophysics and cosmology. 
 

\section{Measurement Overview}
\vspace*{-9pt}



Measuring the $21\,\textrm{cm}$ signature of large scale structure promises enormous scientific payoffs.  Unfortunately, the signal amplitude is small.  These large volume surveys exploit large bandwidths, large fields of view, long integration times, and large numbers of receivers to maximize mapping speed.  Over the past decade, measurements have evolved from using shared-facility single-dish radio telescopes and interferometers to dedicated $21\,\textrm{cm}$ radio interferometers.  Unlike facility interferometers (e.g. SKA),  which include a mix of long and short baselines for high angular resolution imaging, intensity mapping interferometers are close-packed in order to map large-scale cosmological features.  
 The recent revolution in low-noise radio technology allows the construction of relatively inexpensive interferometric arrays with hundreds to thousands of  elements, bandwidth $\gtrsim 100$~MHz, and correlators powerful enough to process the signals from them.  However, R\&D can be carried out efficiently with small demonstrators, with on order ten elements. (See Table \ref{tab:current}).  
 
So far, Cosmic Dawn/EoR experiments have set upper limits on the $21\,\textrm{cm}$ power spectrum\cite{PAPER_Ali_2015,PAPER_Ali_2018}. These limits have steadily improved with better understanding and control of systematic effects.  Post-EoR experiments have detected the $21\,\textrm{cm}$ power spectrum in cross-correlation with galaxy redshift surveys\cite{2010Natur.466..463C, 2013ApJ...763L..20M,2013MNRAS.434L..46S,2018MNRAS.476.3382A}
but none has so far detected it in autocorrelation, i.e. without the assistance of an optical galaxy redshift survey.  These experiments face several measurement challenges, which we describe here. Later we outline how to overcome them with a coherent development plan.

\medskip
\noindent{\bf  Astrophysical foregrounds.}
Astrophysical foregrounds, primarily synchrotron emission from the Galaxy and unresolved point sources, pose the most significant measurement challenge, having much higher
intensity than the $21\,\textrm{cm}$ signal at all frequencies.  These foregrounds have a smooth spectral shape and hence can in principle be distinguished from the $21\,\textrm{cm}$ emission from large scale structure~\cite{2017PASA...34...33P,2013ApJ...768L..36P,2016MNRAS.456.3142S,2012MNRAS.419.3491L}. However, any frequency dependence in the instrument response, for example from the instrument beam or gain fluctuations, can complicate separation of the smooth foreground and the `spikey' $21\,\textrm{cm}$ signal~\cite{Shaw:2014vy,Shaw:2013tb}.  Removing these foregrounds drives most instrument design choices including array element uniformity, cross-polarization response, etc. 

\medskip
\noindent{\bf  Instrument calibration.}
Each antenna has a characteristic response to an input sky signal, known as the instrument gain, which varies with both time and frequency. Each antenna also has a characteristic response on the sky, known as the instrument beam pattern. The gain and beam patterns must be
measured well enough to remove astrophysical foreground
power.  
Even in a `perfect' instrument, the beam varies with frequency and causes features to appear in the spectrum that masquerade as structures in redshift space.  If the frequency-dependent instrument response is known, these foregrounds can, in principle, be subtracted\cite{Ansari2012}.  For actual post-EoR experiments, simulations for CHIME have provided a scale to the problem: the
instrument response on the sky (`beam') must be understood to 0.1\%,
and the time-dependent response of the instrument (`gain') must be
calibrated to 1\% ~\cite{Shaw:2013tb,Shaw:2014vy}.  Uniformity and stability specifications must be carefully integrated into the instrument design and verified during testing and deployment.   


\medskip
\noindent{\bf  FFT beamforming, real time calibration, redundant baselines, and array redundancy.}
For current arrays, with up to about 1000 elements, one can compute the fundamental observables of an interferometer, the correlations  of RF signals from pairs of antennas (visibilities), for all pairs, using existing correlator designs.   Calibration occurs offline.
However,  the correlation cost and data rate from future, larger arrays will
require implementing other correlators, such as the EPIC correlator\cite{Kent2019}, or FFT beamforming correlators. Both 
use FFT-based sampling of the interferometric
geometry~\cite{2004NewA....9..417P,tegmark2009,2010PhRvD..82j3501T,2017arXiv171008591M} to reduce the computational
correlation cost, from order $N^{2}$ to $N\log N$ and output data
volume from $N^{2}$ to order $N$.  Although promising first steps with EPIC and FFT beamforming\cite{Foster2014,Kent2019} have been taken, they have not yet been demonstrated for $21\,\textrm{cm}$ intensity mapping.

FFT beamforming in particular requires that all elements of the array be redundant (that their beams
be similar), placing tight requirements on element uniformity. In
addition, this correlation is performed in real time, and so
requires real-time calibration to account for
instrumental changes (or that the instrument remain extremely
stable). Real-time calibration schemes are being developed \cite{Beardsley2017}. Some exploit the fact that similar interferometric baselines should see the same sky signal and so differences between them can be used to assess relative instrument gains over time. This technique, known as `redundant baseline' calibration~\cite{2010MNRAS.408.1029L,2018arXiv180500953C,2018MNRAS.477.5670D,2014ASInC..13..393R,2016ApJ...826..181D,2017arXiv170101860S}, requires uniform interferometric elements with uniform spacing. And while promising, redundant baseline calibration has not yet been demonstrated to be good enough for $21\,\textrm{cm}$ intensity mapping\cite{Orosz2019,Byrne2019}.



\medskip
\noindent{\bf  Environmental considerations.} In addition to astrophysical foregrounds, two terrestrial contaminants must be eliminated or otherwise mitigated: human-generated radio-frequency interference (RFI) and the ionosphere. Radio bands within the entire range of $21\,\textrm{cm}$ redshifts are popular for communications. RFI appears at
discrete frequencies and can be reduced
or eliminated by a suitable choice of radio-quiet
observation site\cite{2015PASA...32....8O}. 
Even experiments operating in locations with high degrees of interference, notably LOFAR (located in the
Netherlands), have developed impressive RFI removal algorithms
\cite{2012A&A...539A..95O}.

The ionospheric plasma disturbs long wavelength measurement through both refraction and Faraday rotation\cite{2018arXiv180906851T,2017MNRAS.471.3974J}.  The ionosphere acts in concert with the Earth's magnetic field to rotate the polarization vector of incoming light by Faraday rotation. The rotation is proportional to $\lambda^{2}$ as well as the number of free electrons present in the ionosphere, which varies across all time scales. While the $21\,\textrm{cm}$ signal is unpolarized, most foreground emission from the Galaxy is polarized and adds a time-variable component to the foreground characterization and removal. The $\lambda^{2}$ dependence means it primarily affects experiments at longer wavelengths (frequencies below $\sim$ 500\,MHz, $\sim z>2$), which attempt to measure and remove this rotation using accurate maps of the magnetic field and GPS data to infer free electron content. Because signal propagation through the ionosphere is critical for satellite telecommunications, it is well modelled and current low frequency radio telescopes are working to remove signal variability from the ionosphere~\cite{2015PASA...32...29A}. At even lower frequencies ionospheric refraction, which also scales with $\lambda^{2}$, probably becomes the dominant effect .


\medskip
\noindent{\bf  Required sensitivity.}  In the absence of systematic effects, detecting the $21\,\textrm{cm}$ signal requires fielding instruments including thousands of receivers to reach the mean brightness
temperature of the cosmological $21\,\textrm{cm}$ signal of $\sim$ 0.1 - 1\,mK within a few years. Instrument noise stems from a combination of intrinsic amplifier noise (noise temperatures for state-of-the-art radio telescopes range from 25\,K cryogenic to 100\,K uncooled) and sky brightness temperature (which span between
10\,K - 1000\,K depending on pointing and frequency). Because synchrotron emission increases at lower frequencies, at high redshifts
(above $z\sim$ 3) the system noise is dominated by the sky and no
longer by the amplifier; improved sensitivity must be achieved by
fielding more antennas rather than better performing front-end amplifiers. 

\medskip
\noindent{\bf Computing scale.}  Radio astronomy has always been at the forefront of `big data' in
astronomy. Current generation $21\,\textrm{cm}$ instruments produce $\gtrsim 100$\,TB of
data per day without any compression.  The data volume $\propto N^{2}$ where $N$ is the number of elements (currently $N\sim 10^{3}$),
representing challenges in data reduction, transfer, storage, analysis, distribution, and simulation. Compression by a factor of $\sim N$ is achievable with the FFT beamforming mentioned above. 
To aid in data transport, analysis, and data quality assessment, data must be compressed further (e.g. co-adding maps in a weekly cadence). This reduces the data size but increases pressure on real-time instrument calibration. In addition, to enable transient science,  fast triggers are deployed at some current generation instruments\cite{CHIME2019}. 

\section{Technology/hardware Development}
\vspace*{-9pt}

Some of these challenges require advances in instrumentation. 
The aim is to have interferometers made of a large number of elements, each one reasonably inexpensive and robust, ensuring easy integration, and smooth operation. 

\noindent{\bf Early digitization and signal processing.}
\label{Electronics}
   As noted above, changes over time of the gain of the receiver electronics is one of the limiting factors in removal of astrophysical foreground power.  One promising idea to improve stability is to digitize the analog signal directly at each antenna, rather than remotely, which is the current practice.   This early digitization will avoid the use of long analog cables, and analog frequency downconversion.  This approach requires stable analog amplifier systems at each antenna and avoidance of RFI generated by the digital electronics.  Furthermore, the system clock signal must be distributed precisely to each of the digitizers, moving the instrument phase calibration problem from the
analog system into the clock distribution system.  Thermal stabilization of the low noise amplifiers (LNAs) at each antenna may be required.



\noindent{\bf Optical and analog design.}
The receiver noise temperature is dominated by loss in the antenna as well as the noise in the LNA. HIRAX, for example,  has chosen to reduce the system noise by up to 30\% by fabricating the LNA directly in the antenna itself, reducing the transmission loss and taking full advantage of low-noise transistors available in these bands. 


Antennas for $21\,\textrm{cm}$ experiments range from dipoles for long wavelength experiments to parabolic reflectors, including on-axis dishes as well as cylindrical designs, at shorter wavelengths.    Low sidelobes are desirable to allow operation when the Sun is up and to minimize coupling chromatic response into foregrounds. 
So far, the parabolic reflectors have supported receivers at the focus with struts or tensioned cables, leading to some diffraction and reflections. To illuminate the reflector, they have also included variants of dipole feed antennas with wide beams that have non-negligible cross-talk and frequency-dependence.   These choices are typically made to save cost and complexity, but make calibration more difficult. Further studies should include options such as off-axis geometries (like SKA-mid and ngVLA) and possibly horn/Gregorian receivers, keeping marginal costs low while meeting uniformity and  bandwidth flatness specifications, and exploring new reflector fabrication techniques. Current instruments are limited by the fact that their interferometric elements are not similar enough and motivate assessment of fabrication tolerances in mass-production.  
 


\noindent{\bf Calibration.}
\label{beam_characterization}
Current instruments rely primarily on sky signals for calibration of the beam and gain.  However, this approach has not yet been demonstrated to adequately remove foregrounds with these instruments. The frequency-dependent gain for each input must be known to $\sim$ 1\% on time scales between the integration period ($<5$\,s scales) and a few hours, depending on the rate of appearance of on-sky radio calibration sources~\cite{Shaw:2014vy}. This challenge can be met by a combination of instrument stability and a suitable calibration scheme. CHIME~\cite{2014SPIE.9145E..22B,2014SPIE.9145E..4VN} is updating a classic radio noise-injection scheme which can be used to calibrate many signal chains at once. To implement such an active calibration technique for large arrays will require development of a stable calibration distribution network as well as passive models of gain and beam variation with temperature and dish pointing. 

 Because the antenna beam pattern (main beam, sidelobes, and polarization) is frequency-dependent, it can mix frequency dependence (related to redshift) and sky location.  This problem is expected to be the primary source of contamination from foreground emission into the signal band, and so must be known even more accurately than the electronic gain ($\sim$0.1\%)~\cite{Shaw:2014vy}. This level of calibration is difficult for $21\,\textrm{cm}$ telescopes because they are stationary and designed to have large beams for improved survey speed \cite{2017PASA...34...62S}. In addition, some instruments (such as CHIME) have large antennas, which can be difficult to simulate. 
 Many $21\,\textrm{cm}$ instruments are beginning to use signals from small unmanned aerial systems (sUAS, or ``drones") and satellites 
 to map the beam shape and measure the gain\cite{Jacobs_drones, neben:2016,Virone_drones, Pupillo_drones,Chang_drones,Line:2018}. 
 Ultimately, the beam calibration requirement sets a specification on uniformity in antenna fabrication.

\begin{figure}[htb!]
  \centering
  \includegraphics[width=5in]{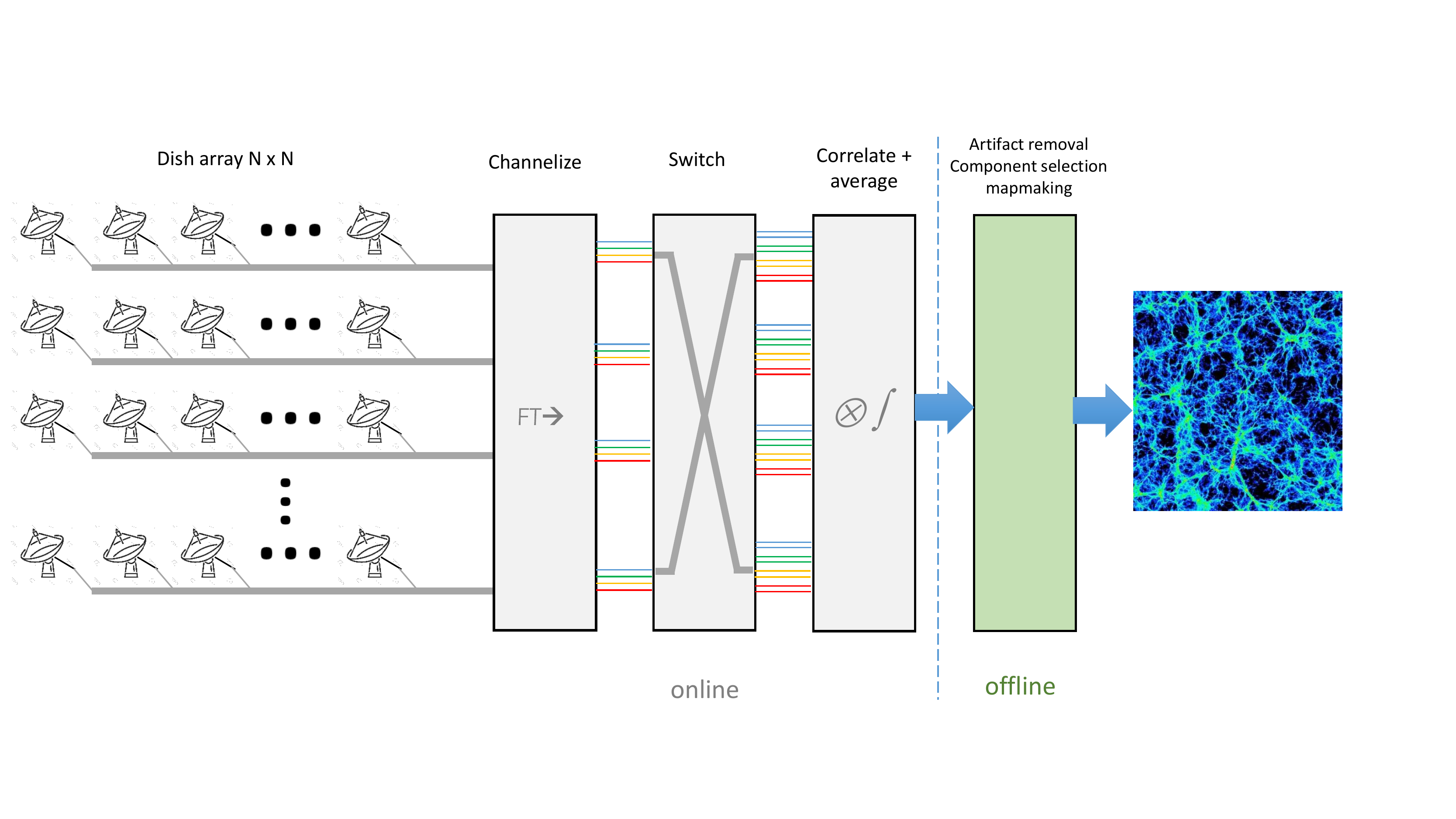}
\caption{\footnotesize Illustration of anticipated data flow in a large interferometric array.  Conversion of waveform data to frequency space, e.g. channelization, is accomplished close to each receiver; coincident data for each frequency bin are collected from all stations through a cross-bar switch (also called a ``corner-turn'' operation); correlations are constructed for each frequency bin, which can then be time-averaged and stored, followed by physics analysis.}
\label{fig:data_flow_block_diagram}
\end{figure}

\noindent{\bf Realtime data flow and processing.}
\label{data_flow}
 Large-N arrays are becoming increasingly software based, particularly when it comes to real-time data compression.  Computing requirements for large interferometers come from the correlation burden, real-time calibration, RFI-excision, and the data reduction, transfer, storage, analysis, and synthetic data production (Fig.\ref{fig:data_flow_block_diagram}). 
 The correlator computation requires development in computing approaches which can improve the cost scaling both for equipment and power. Examples could include using commodity-hosted FPGA's, combined FPGA/CPU systems\cite{Intel1,Intel2}, using/developing dedicated ASIC's \cite{7436238}, or using GPUs to exploit fast-paced hardware updates for correlator computation.


\section{Data Analysis Development}
\label{sec:data_analysis}

Releasing science deliverables for the community from a $21\,\textrm{cm}$ experiment depends crucially on developing an analysis pipeline that can transform vast quantities of data into well characterized frequency maps and power spectra. This is a computationally costly exercise, but does not require continuous real time processing. The analysis challenges center both on the methodology (optimal algorithms have yet to be demonstrated) and software engineering (data volumes are incredibly large). We can divide the analysis up into three broad areas.

\noindent {\bf Flagging, calibration, and pre-processing at scale.} The acquired data are processed to reduce remaining systematic effects.
Of particular importance is cleaning of RFI by flagging times and frequency channels that have been contaminated. The problem is well understood within radio astronomy \cite{2012A&A...539A..95O}, though the effects of residual RFI at the small level of the $21\,\textrm{cm}$ signal is only starting to be addressed \cite{2018MNRAS.479.2024H}.
Though much of the calibration must be done in real-time (see above) there are still degrees of freedom that must be corrected in later analysis. 

\noindent {\bf Astrophysical foreground removal.} 
Foreground mitigation falls broadly into two classes: foreground
avoidance and foreground cleaning. Foreground avoidance is the
simplest of these two approaches, relying on the fact that
contamination produced by a typical interferometer configuration is strongest in certain regions of $k$-space. Producing cosmological results using only the cleanest modes is a simple and effective technique, but it becomes deeply unsatisfactory at low frequencies, particularly in the dark ages. Here Galactic synchrotron and extragalactic point source radiation quickly becomes very bright 
even at high Galactic latitudes. At the same time, the window of clean modes dramatically narrows due to the relative scaling of the angular diameter distance and Hubble parameter with redshift \cite{2015MNRAS.447.1705P}. Combined, this means that at a given threshold for contamination we exclude increasingly large regions of $k$-space at high redshifts, significantly degrading any cosmological result.

Foreground cleaning instead of (or in conjunction with) foreground avoidance then becomes an attractive option. These  methods rely on detailed knowledge of the instrument response and the sky 
to predict and subtract the actual foreground signal. 
The residual contamination is set by both the amplitude of the raw contamination and the accuracy with which the beam has been measured. 

\noindent {\bf Cosmological processing.} The next step after foreground cleaning is to compute cosmologically useful quantities such as power spectra and sky maps.  This step has been done within the CMB and LSS communities for many years but radio interferometric data brings unique challenges. Nevertheless, conceptual frameworks have been developed to tackle these problems\cite{Shaw:2014vy,2014PhRvD..90b3018L,2004ApJ...609....1J}.  
Several areas of data analysis, then, will require research investment to ensure the success of a large scale $21\,\textrm{cm}$ intensity mapping survey:

\begin{enumerate}

\item{\bf Scaling.}  A significant challenge is scaling existing analysis techniques to work with the vast increase in expected data in an energy-constrained/post-Moore's Law computing landscape. This process will require optimization of algorithms to reduce the computational cost, and ensuring that the techniques can operate in parallel on leading edge supercomputers.

\item{\bf Systematic robustness.} Both astrophysical uncertainties (such as the exact nature of foregrounds) and instrumental uncertainties (such as calibration and beam optics) cause foreground contamination. Developing more robust cleaning techniques will reduce systematic biases, and may lead to cost savings by reducing instrumental tolerance requirements.

\item{\bf Improving signal recovery.}  The loss of significant numbers of modes during foreground removal reduces our constraining ability generally and particularly affects science that needs access to the largest scales. Foreground removal methods like tidal reconstruction \cite{Zhu:2015zlh,Foreman:2018gnv} need to be developed that reduce the effect of the `foreground wedge'\cite{Dillon2014}. Similarly, traditional reconstruction techniques \cite{2007ApJ...664..675E,2009PhRvD..79f3523P} that recover non-linear modes need to be adapted for the peculiarities of $21\,\textrm{cm}$ intensity mapping.

\end{enumerate}

\section{Simulation needs \& challenges}
\label{sec:simulation_challenge}

The challenges facing $21\,\textrm{cm}$ surveys are significant but well understood. However, our ability to tackle them requires sophisticated approaches in instrumental design and offline analysis. It is therefore essential to use simulations to close a feedback loop that allows prediction, and thus refinement, of the effectiveness of a design and analysis strategy. The importance of this testing is especially evident in the face of sophisticated foreground mitigation strategies with the potential for significant signal loss \cite{Cheng:2018}.
While expensive to implement, an end-to-end simulation of the entire measurement process is necessary for designing a successful instrument and observational program.

Producing realistic simulations of data from any instrument configuration and propagating these to final cosmological results is a conceptually straightforward prospect:
\begin{enumerate}
  \item{\bf Produce a suite of full-sky maps of the ``true'' sky.} These include both the $21\,\textrm{cm}$ cosmic signal and foregrounds, with one map per frequency bin observed by the instrument. 
  
  \item{\bf ``Observe'' these maps with a simulation pipeline.} 
  This step requires a detailed enough instrument model to provide realistic mock visibility data streams, including effects such as far side lobes, instrumental parameters drift, etc.; and 
  
  \item{\bf Feed these mock observations into the data analysis pipeline and produce reduced data cosmological analyses.} This pipeline, discussed Section \ref{sec:data_analysis}, is the same that would be used on real data. This step requires software tools to compute power spectra ($P(k)/C(\ell,\nu)$) from 3D sky maps or directly from visibilities, or higher order statistics. These tools could then be used to evaluate the realistic performance of different instrument configurations for constraining cosmological models and parameters.
\end{enumerate}

\begin{figure}[htb!]
  \centering
  \includegraphics[width=0.44\linewidth]{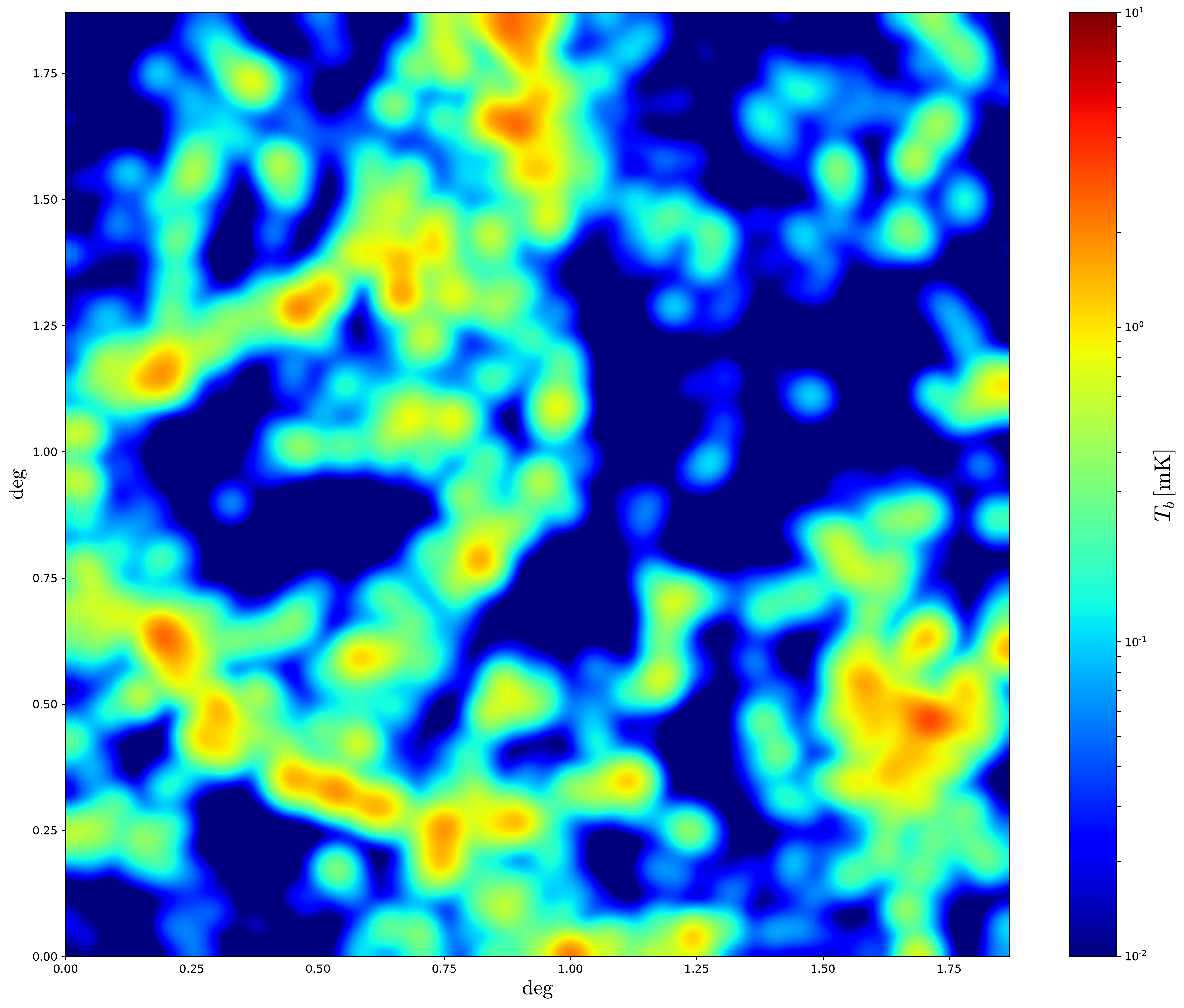}
  \includegraphics[width=0.44\linewidth]{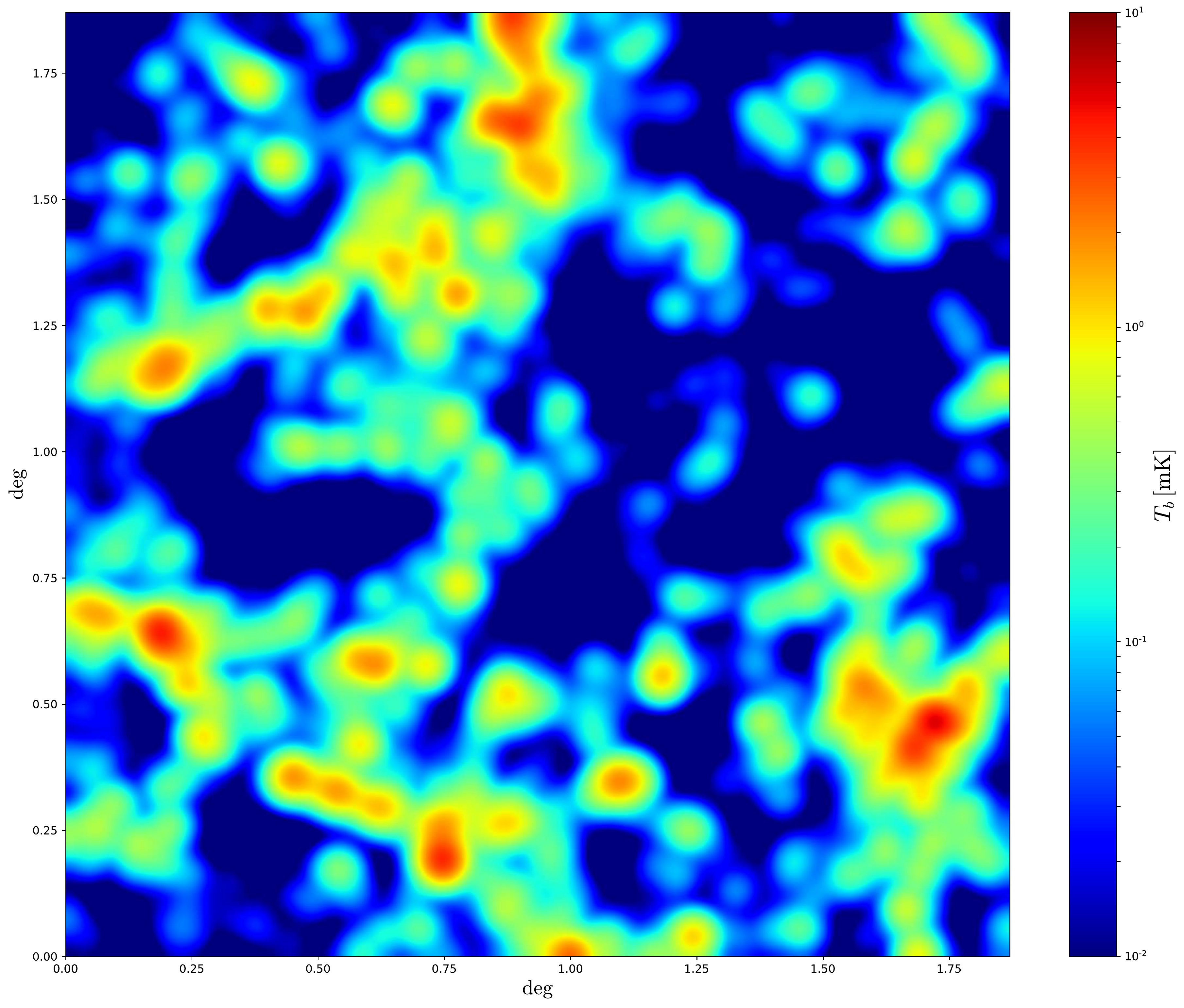}
  \caption{\footnotesize $21\,\textrm{cm}$ maps at a frequency of 710~MHz over a channel width of 1~MHz with an angular resolution of 1.5' over an area of $\simeq4~{\rm deg}^2$. The map on the left has been created from the state-of-the-art magneto-hydrodynamic simulation IllustrisTNG with a computational cost of $\simeq$18 million cpu hours. The map on the right panel has been generated by assigning HI to dark matter halos of an N-body simulation using the a simplification of the ingredients outlined in \cite{2018ApJ...866..135V}. The computational cost of the N-body simulation is much lower than that of the full hydrodynamical simulation, and allows modelling of the HI field in a very precise and robust manner. The shapes of the power spectra of these two simulations differ by only $\sim 5$\% from the largest scales down to $k = 1 {\rm h~Mpc}^{-1}$.  
  }
  \label{fig:21cm_sim}
\end{figure}

For verification of foreground removal effectiveness, Gaussian or pseudo-Gaussian $21\,\textrm{cm}$ simulations are largely sufficient \cite{2014MNRAS.444.3183A,Shaw:2014vy}. However, for targeting sensitivity to specific effects (e.g. non-Gaussian initial conditions), or in cross-correlation with other probes, more accurate simulations constructed from mock-catalogues will be required. This allows us to produce correctly correlated maps for additional tracers (e.g. LSST photometric galaxies), and also for radio point source contribution to the foregrounds.

Though the relation between HI density and total matter density involves complex environment-dependent processes, simulating it can be done efficiently. Recent work has shown that one can take advantage of the fact that neutral hydrogen in the post-reionization era resides almost entirely inside dark matter halos~\cite{2018ApJ...866..135V}. 
Thus, one can calibrate the relation between dark matter halos and HI using hydrodynamic simulations and create $21\,\textrm{cm}$ maps (Fig. \ref{fig:21cm_sim}) via less expensive methods such as N-body or fast numerical simulations\cite{Pinocchio,ALPT,Halogen,EZMock,PATCHY,COLA,QuickPM,FastPM,PeakPatch}.  
Similarly, while there are large-scale hydrodynamic EoR simulations (e.g. DRAGONS), modelers are parameterizing these numerical methods, and finding efficient ways to constrain those parameters \cite{2018IAUS..333...18G}.


Galactic synchrotron is the largest foreground contaminant and simulations must ensure they are not artificially easy to clean. A simple approximation can be produced by proceeding from a full sky map at a radio frequency 
and scaling to different frequencies based on the known spectral index of Galactic synchrotron radiation. However this is not sufficient at the dynamic range between the foregrounds and the $21\,\textrm{cm}$ signal and one must be careful to include: spectral variations about a pure power law; small scale angular fluctuations not captured in existing surveys; and polarization, including the effects of emission at a wide range of Faraday depths which generates significant spectral structure in the polarized emission \cite{Shaw:2014vy}. More sophisticated Galactic models, for example from MHD simulations, could also be developed and used here.  Additional observations with existing instruments will better characterize the spectral behavior of the foregrounds.

In Step 2, a realistic instrument simulation pipeline would take the maps discussed and convolve them with the complex beam for each
antenna in the interferometer\cite{pyuvsim}. 
This can be done by direct convolution
utilizing the fact that for a transit telescope it is sufficient to
generate a single day of data. However, for wide-field transit
interferometers this can be performed more efficiently in harmonic
space using the $m$-mode formalism\cite{Shaw:2013tb} ($O(N\log{N})$ instead of
$O(N^2)$). Some of the required code would be similar to code used in CMB science, such as
the TOAST
package 
using fast numerical techniques for beam convolution
\cite{2010ApJS..190..267P}.

These simulations require realistic models of the telescope beams. Electromagnetic simulation codes such as CST, GRASP and HFSS can be used for this, but achieving the accuracy required is challenging computationally \cite{2017JAI.....650003C,2017arXiv170808521D,2016ApJ...831..196E}. A complementary approach is to generate synthetic beams with sufficient complexity to capture the challenges posed by real beams.   These are computationally easier to produce, but must be informed by real measurements and electromagnetic simulations to ensure their realism, and may be aided by machine learning algorithms.

\begin{table}[htb!]
  \centering
  \resizebox{\textwidth}{!}{
  \begin{tabular}{lccccrc}
    Name & Optimized & Steerable & Type  & Elements & Redshift & First light \\
    \hline \\    
    \multicolumn{1}{p{2.9cm}}{\uline{Existing w/ data:}} \\
    GMRT~\cite{2013MNRAS.433..639P} & N & Y & interferometer & 30 dual-pol\,$\times$\,45\,m dishes & 28 & 1995\\
    GBT\cite{GBT_Switzer_2013} & N & Y & single dish & 1\,dual-pol on $100\,$m dish & $\sim$0.8 & 2009  \\
\\    \multicolumn{1}{p{3.3cm}}{\uline{Dedicated exp'ts:}} \\

    OVRO-LWA\cite{LWA_Eastwood_2018,LWAnote} & Y & N & interferometer &  288 dual-pol dipoles & 16 -- 52 & 2012 \\
    
    MWA\cite{2013PASA...30....7T,MWAnote} &  Y & electronic & interferometer & 256 tiles of 16 dual-pol dipoles & 3.7--16 & 2013 \\
    
    PAPER\cite{2010AJ....139.1468P,PAPERnote}  & Y & N & interferometer & 128 dual-pol dipole/reflector & 7--12 & 2010 \\
    
    HERA\cite{2017PASP..129d5001D,HERAnote}  &  Y & N & interferometer & 350 dual-pol 14~m dish (staged deploy't) & 6--30 & 2016\\
    
    LOFAR\cite{2013A&A...556A...2V,LOFARnote} & Y & electronic  & interferometer & 3400 tiles of 16 dual-pol dipoles & 5--11 & 2007 \\
    
     &  &  &  & 4224 dual-pol dipoles & 17--141 & \\
    
    CHIME\cite{CHIME,CHIMEnote} & Y & N & cylinder interferometer & 1024 dual-pol over 4 cyl   & 0.75 -- 2.5 & 2017  \\
    HIRAX\cite{HIRAX,HIRAXnote} & Y & limited & dish interferometer & 1024 dual-pol\,$\times$\,$6\,$m dishes
                                                & 0.75 -- 2.5 & 2020  \\
    Tianlai Dish\cite{Tianlai,Tianlainote} & Y & Y & dish interferometer & 16 dual-pol\,$\times$\,$6\,$m dishes& 0 -- 1.5 & 2016 \\
    Tianlai Cylinder\cite{Tianlai,Tianlainote} & Y & N & cylinder interferometer & 96 dual-pol
                                                         over 3 cyl   & 0 -- 1.5 & 2016 \\
                                                         
    OWFA\cite{OWFA_Chatterjee_2018,OWFAnote} & N & Y & cylinder interferometer &  264 single-pol& $\sim$ 3.4$\pm$0.3 & 2019 \\
    
    BINGO\cite{BINGO,BINGOnote} & Y & N & single dish  & $\sim$60 dual-pol sharing 
                                   $\sim$50\,m dish & 0.12 -- 0.45 & 2020 \\
\\    \multicolumn{1}{p{2.9cm}}{\uline{Dedicated R\&D:}} \\
    BMX & Y & N & dish interferometer & 4 dual-pol\,$\times$4\,m
                                             off-axis dishes & 0 -- 0.3 & 2017 \\
    NCLE\cite{NCLE_Flacke_2018} & Y & N & satellite  & 3$\times$5\,m monopole ant. at
                                Earth-Moon $L_2$ & $>17$ &  2018\\
                                
    PAON-4\cite{PAON4_Zhang_2016,PAON4note} & Y & limited & dish interferometer & 4 dual-pol\,$\times$5\,m dishes & 0 -- 0.14 & 2015 \\    
    
\\    \multicolumn{1}{p{2.5cm}}{\uline{Non-dedicated:}} \\

    MeerKAT & N & Y & single-sish & 64 dual-pol\,$\times$\,13.5\,m dishes & 0 -- 1.4 & 2016 \\
    
    SKA1-MID & N & Y & single-dish & $\sim 200$ dual-pol\,$\times\sim$\,15\,m dishes & 0 -- 3 & 2028 \\
    
    SKA1-LOW\cite{SKA-LOWnote} & N & electronic & interferometer & 269,312 dual-pol dipoles & 3--27 & 2028\\
\\    \multicolumn{1}{p{2.5cm}}{\uline{Proposed:}} \\
    PUMA\cite{Ansari2018,PUMAWhitePaper} & Y & limited &  dish interferometer & 5000, 32000 or 64000
                                                                dual-pol\,$\times$\,
                                                               6\,m dishes & 0.3 -- 6 & $<$2030 \\
  \end{tabular}
 }
  
  \caption{\footnotesize Current and planned $21\,\textrm{cm}$ intensity mapping experiments.  The ``First
    light'' column refers to first light for $21\,\textrm{cm}$ observations for
    non-dedicated experiments. In the ``Optimized'' column, we note
    whether the telescope has been designed with intensity mapping as
    its primary scientific goal. 
    For MeerKAT and SKA-MID,
    dishes will likely be used in a single-dish mode, with
    interferometric capability used only for gain calibration.
    }
  \label{tab:current}
\end{table}

Capturing non-idealities in the analog system, particularly gain variations, is mostly straightforward as these can be applied directly to the ideal timestreams. Additionally, one needs to include time-dependent beam convolution (including position and brightness) for temporally varying sources such as solar, jovian and lunar emission as well as the effects of RFI at low levels~\cite{2018MNRAS.479.2024H}.
Including calibration uncertainties poses a particular challenge, because of the realtime calibration and compression of the instrument.  Simulating these effects requires either: generating data at the full uncompressed rate, applying gain variations, and then performing the calibration and compression processes; or the computationally easier alternative of generating models of the effective calibration uncertainties.

After the first two stages, mock observations are then fed to the proposed data analysis pipeline, and propagated through to final cosmological products, to assess analysis systematics, instrument design, real-time calibration, etc. to determine whether the pipeline is sufficient to meet the science goals. Though the simulation program is well defined, there are many open challenges to address:

\begin{enumerate}

\item{\bf Understanding the HI Distribution.} To map the HI distribution to the cosmologically useful matter distribution requires cutting edge hydrodynamic simulations to capture the small halos that HI favours over a cosmologically interesting volume. 

\item{\bf Scale.} 
We need to be able to produce large numbers of emulators that Monte-Carlo over the experimental uncertainties. 

\item{\bf Improving the feedback loop.} While a straightforward version of the simulation loop above can tell us whether a proposed design meets requirements, it does not show how to improve the design to ensure that it does. For a complex instrument with many design parameters it is essential to be able to guide this process to find the most relevant combinations of changes.

\end{enumerate}

 \section{Participation in Current and Near-term (Stage I) Programs}

Support for both small dedicated test-bed instruments and participation in existing and future large-scale initiatives, both US-led and international,  will enable development and testing of the methods presented above and will benefit $21\,\textrm{cm}$ intensity mapping efforts ranging from the Dark Ages to the post-EoR epoch.
We believe that with rather modest investment of resources, such R\&D will enable current experiments to extract maximum science while paving the way for future and even more exciting telescopes (Table \ref{tab:current}). In particular, these efforts are the only way to make progress in foreground removal, instrument calibration, correlating signals from large arrays, dealing with RFI, achieving adequate sensitivity, and analyzing huge data cubes.  

While there has been considerable funding from the US for Cosmic Dawn/EoR programs, currently there is no significant US funding for any of the existing post-EoR Stage I programs (CHIME, HIRAX, Tianlai). Post-EoR $21\,\textrm{cm}$ intensity mapping is an unexploited probe of dark energy and inflation that is complementary to CMB and optical surveys.  Participation in these efforts, or a new effort, would enable a coherent development plan that would include:
\begin{enumerate}
\item Construction of a test bed with 10-100 interferometer elements to test technologies (antennas and electronic as well as internal calibration systems).
\item Construction of a large, $\sim 1000$ element array to understand specific problems of operating such an array, including calibration and stability issues. Such an array should be operated with a (pairwise) correlator system first.
\item Development of an FFT beamformer and associated real-time calibration for this array. 
\item Construction of a Stage II array with $10^4 - 10^5$ elements.  An instrument of this scale is necessary to survey the entire post-EoR epoch.
\end{enumerate}

Hardware development is not enough. An increasingly large share of the budget for all $21\,\textrm{cm}$ intensity mapping programs is in the development of production- or system-level software.  We need to pay close attention to developing software that does the real-time analysis (calibration, RFI excision, etc.), integrate it into a system, and fully test it.



\section{Conclusion}

A wide range of Astro2020 science goals depend on the success of $21\,\textrm{cm}$ intensity mapping for measuring large scale structure in the universe over enormously larger volumes than possible today.   This white paper has described a plan to advance this technique over the next decade and includes 1) development and testing of key technologies for reliable, robust single antenna systems which could be assembled as transit interferometric arrays, and associated central real time control/acquisition and computing systems.  Most subsystems would be common to Cosmic Dawn/EoR and post-EoR observations.  It also includes development of software for 2) data analysis and 3) simulations to realistically evaluate different instrument configurations. 
A significant fraction of the software tools would be common to all epochs. These technical achievements require 4) support for participation in current (Stage I) experiments to lay the groundwork for next generation (Stage II) experiments.  As described above, Stage I experiments have already taught many lessons which must be understood before embarking on the next generation.

\clearpage

\bibliographystyle{unsrt}
\bibliography{main}

\end{document}